\documentclass[twoside,a4paper,11pt]{sca}
\pdfoutput=1
\usepackage{graphicx}
\usepackage{hyperref}
\usepackage{movie15}
\usepackage{natbib}  
\usepackage{epstopdf}

\def\la{\mathrel{\mathchoice {\vcenter{\offinterlineskip\halign{\hfil
$\displaystyle##$\hfil\cr<\cr\sim\cr}}}
{\vcenter{\offinterlineskip\halign{\hfil$\textstyle##$\hfil\cr
<\cr\sim\cr}}}
{\vcenter{\offinterlineskip\halign{\hfil$\scriptstyle##$\hfil\cr
<\cr\sim\cr}}}
{\vcenter{\offinterlineskip\halign{\hfil$\scriptscriptstyle##$\hfil\cr
<\cr\sim\cr}}}}}

\begin{document}
\pagenumbering{arabic}
\pagestyle{myheadings}
\thispagestyle{empty}
{\flushright\includegraphics[width=\textwidth,bb=90 650 520 700]{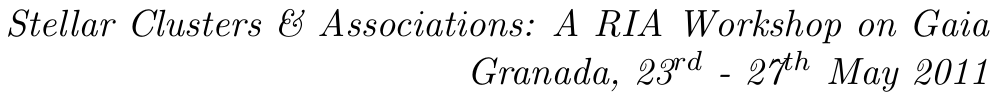}}
\vspace*{0.2cm}
\begin{flushleft}
{\bf {\LARGE
%
Young star clusters in external galaxies
%
}\\
\vspace*{1cm}
%
S{\o}ren S. Larsen
%
}\\
\vspace*{0.5cm}
%
Astronomical Institute, Utrecht University, The Netherlands
%
\end{flushleft}
%
\markboth{
Young star clusters in external galaxies
}{ 
%
S{\o}ren S. Larsen
%
}
\thispagestyle{empty}
\vspace*{0.4cm}
\begin{minipage}[l]{0.09\textwidth}
\ 
\end{minipage}
\begin{minipage}[r]{0.9\textwidth}
\vspace{1cm}
\section*{Abstract}{\small
%
I review the characteristics of cluster populations in other galaxies, with particular emphasis on young star clusters and a comparison with the (known) open cluster population of the Milky Way. Young globular cluster-like (compact, massive) objects can still form at the present epoch, even in relatively quiescent spiral discs, as well as starbursts. Comparison with other nearby spiral galaxies, like M83 and NGC 6946, suggests that the Milky Way should host about 20 clusters with masses above $10^5$ M$_\odot$  and ages younger than about 200 Myr. No such clusters have been found, however. I discuss the important roles of selection and evolutionary effects that may account for many of the apparent differences between cluster populations in different galaxies. One potentially important difference between ancient GCs and young star clusters is the presence of complex star formation / chemical enrichment histories in the GCs. Little is currently known about the presence or absence of such features in massive ($>10^5 M_\odot$) young star clusters, but some tantalizing hints of extended star formation histories are now emerging also in young clusters. 
%
\normalsize}
\end{minipage}
%
%
%
\section{Introduction \label{intro}}

The classification of star clusters as either ``open'' or ``globular'' may first have been explicitly adopted by \citet{Shapley1916}. However, it is clear that neither the ``classical'' globular nor open clusters constitute homogeneous classes of objects, and furthermore the boundary between the two classes is somewhat fuzzy. The Galactic globular clusters (GCs) can be grouped into at least two sub-populations, loosely associated with the Galactic halo and bulge/thick disc, that have distinct kinematics, spatial distributions and metallicities \citep{Zinn1985,Minniti1996}. Even in our Galaxy, the open clusters span large ranges in age, mass and concentration that overlap somewhat with the globular clusters, and making a clear-cut distinction between open and globular clusters becomes increasingly difficult as one moves to external galaxies \citep[e.g.][]{Hodge1988}. This has become particularly true in the last few decades due to the discovery of young star clusters with masses in the range $10^5 - 10^7 M_\odot$, predominantly in interacting galaxies and starbursts \citep{Whitmore2003,PortegiesZwart2010}, but also in some normal non-interacting spirals \citep{Larsen2000a}. To accommodate these objects, terms such as young massive clusters (YMCs), super-star clusters (SSCs), etc., have been introduced.

Although classification is often a natural first step in the study of a new field, there is no generally accepted classification scheme for star clusters that has managed to gain foothold, and it is not clear to what extent the different labels that have been adopted by various authors correspond to physically different objects \citep[see an excellent discussion of this point in][]{Terlevich2004}. It therefore appears more useful to discuss the differences and similarities among cluster systems in different environments in terms of quantifiable parameters. To this end, it is useful to distinguish between three different aspects of a cluster system: 1) The formation efficiency of (long-lived) star clusters relative to field stars, 2) the shape of the initial cluster mass function (ICMF), and 3) dynamical effects. Items (1) and (3) are dealt with extensively elsewhere in these proceedings (e.g.\ Bastian, Bressert, Gieles, Kruijssen). Here I will mainly concentrate on (2), the ICMF.

\section{The initial cluster mass function (ICMF)}

The ICMF is analogous to the stellar IMF, and constraining it is subject to many of the same difficulties. Like stars, clusters have finite lifetimes that must be taken into account when trying to infer the ICMF from the observed mass distribution of a cluster sample. 
It is straight forward to show that if clusters lose mass at a rate $\dot{M} \equiv -M / t_{\rm dis}$, with $t_{\rm dis} \propto M^\gamma$, then $dN/dM \sim M^{\gamma-1}$ independently of the ICMF for $t \gg t_{\rm dis}$. Hence, memory of the initial shape of the MF is lost below some mass limit.
This is consistent with observations showing that the present-day MF of ancient globular clusters is about flat at $M \la 10^5 M_\odot$ if $t_{\rm dis}$ scales roughly linearly with cluster mass \citep{Fall2001,McLaughlin2008,Kruijssen2009}. At younger ages, only the low-mass end of the MF is expected to have been strongly affected by dynamical evolution, although the detailed relation between cluster mass and $t_{\rm dis}$ is more uncertain at younger ages (e.g.\ Bastian, these proceedings). At high masses, the main empirical difficulty is that the most massive clusters, like the most massive stars, are rare, and as a result there is still considerable uncertainty about the shape and degree of universality of the high-mass end of the ICMF. Care must be taken to separate statistical size-of-sample effects from real physical differences.

It is worth reflecting for a moment on the exact meaning of the term \emph{initial} cluster mass function. The ICMF may be understood as the mass function ``at birth'' for a cluster population, but this is not a very useful definition in practice. The moment of birth is not well defined, and at young ages the identification of individual star clusters is often difficult (e.g.\ Bressert, these proceedings). In addition, since the most massive clusters are rare, a cluster sample spanning a very small range of ages will not be very suitable for constraining the behaviour of the ICMF at the high-mass end.
A more practical approach is to view the ICMF as a \emph{mean birthrate function}, 
\begin{equation}
  \Psi(M) \equiv \frac{d^2 N}{dM \, d\tau}
\end{equation}
i.e., $\Psi(M)$ is the average number of clusters born per unit mass per unit time interval. This may then be integrated over time and mass in order to yield, e.g., the expected number of clusters born with a given range of masses over a given time interval. 

Several studies have constrained the shape of the ICMF over some mass interval. In the Milky Way, the shape of the ICMF is well approximated by a power-law $dN/dM \propto M^{-2}$ for $\la 10^3 M_\odot$ \citep{Elmegreen1997,Lada2003,Selman2008}. At higher masses the ICMF in the Milky Way is poorly constrained, but the discovery of several young clusters with masses in the range $10^4 - 10^5 M_\odot$ shows that our Galaxy is certainly still capable of forming such objects \citep{Clark2005,Figer2006,Davies2007}. The $M^{-2}$ power-law behaviour has been found in other galaxies too, like the LMC \citep{Hunter2003,DeGrijs2006,Chandar2010}, M51 \citep{Bik2003,Chandar2011} and the Antennae \citep{Zhang1999,Fall2009}, for higher masses than in the Milky Way.

\subsection{Range of validity for the $M^{-2}$ ICMF}

\begin{figure}
\center
\includegraphics[width=9cm]{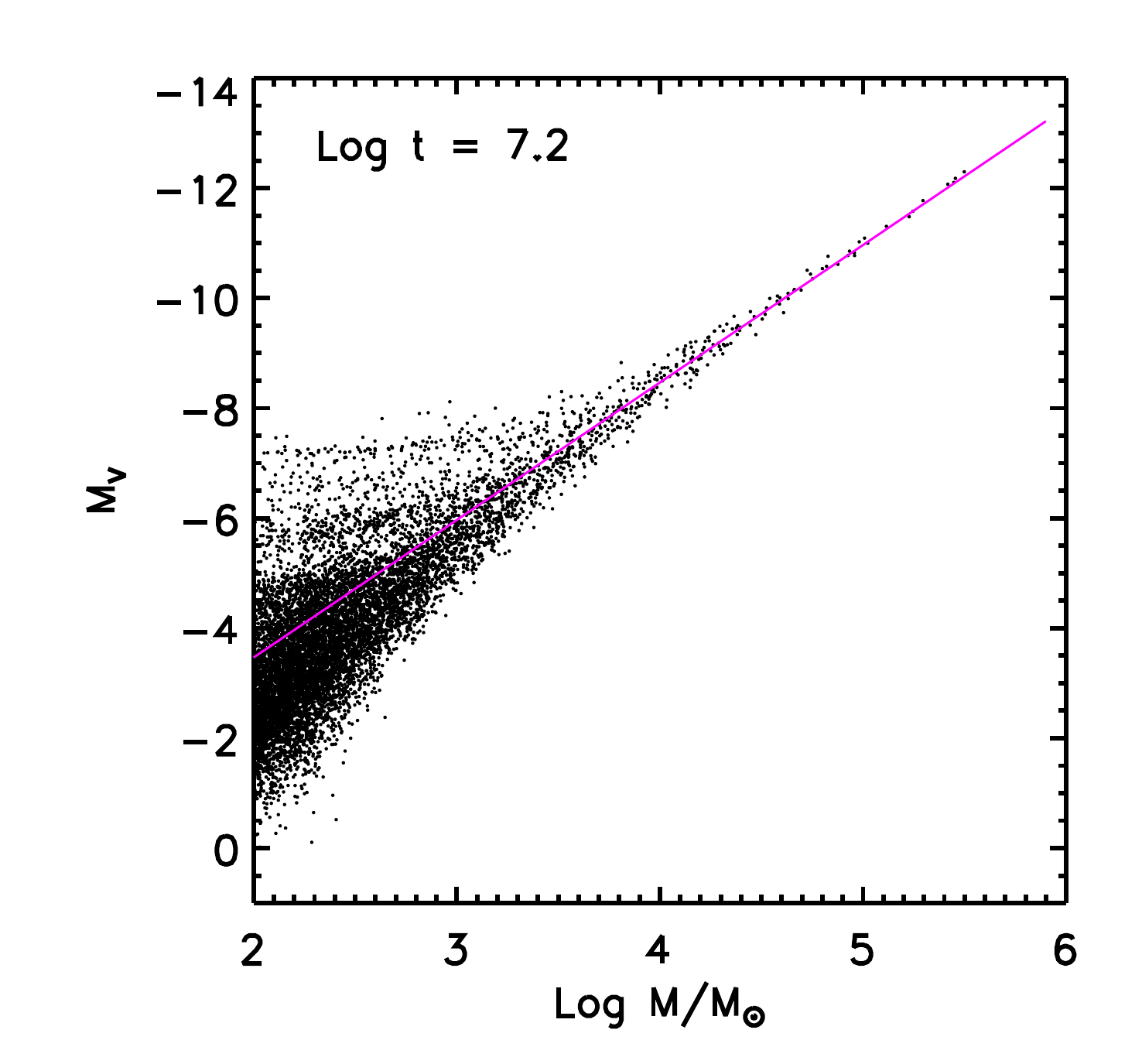} 
\caption{\label{fig1}Absolute $M_V$ magnitude as a function of mass for simulated star clusters with an age of $\log t/yr = 7.2$. Note that there is no unique mapping between $M_V$ and masses less than $\log M/M_\odot \sim 4$, even for fixed age.}
\end{figure}

While the $M^{-2}$ power-law form seems relatively well established for masses up to $\sim10^5 M_\odot$, the upper and lower limits remain less well constrained. This has to do with statistics in both cases, but of quite different nature. At the low-mass end, the discreteness of the \emph{stellar} IMF becomes an important limiting factor especially when relying on the integrated properties of clusters to determine ages and masses, as is typically the case in extragalactic studies. The light from a $10^4 M_\odot$ cluster is dominated by \emph{on average} only 2--3 supergiants for ages up to several tens of Myr. Since this is an average, some clusters will have no supergiants at all, while others will have more than the average value. This leads to very large fluctuations in the integrated magnitudes colours that are \emph{not} symmetrically distributed around the values predicted for a continuously sampled IMF \citep{Barbaro1977,Bruzual2002,Girardi1995,Cervino2004,Cervino2006,MaizApellaniz2009,Piskunov2009,Fouesneau2010,Popescu2010b}. As an illustration of this, Fig.~\ref{fig1} shows the $M_V$ magnitude vs.\ stellar mass for simulated star clusters with an age of $\log t/yr = 7.2$. At low masses there is a very large scatter in this relation, and it only starts to narrow down above $\sim10^4 M_\odot$.
This discreteness of the IMF also affects cluster detection itself; a low-mass cluster dominated by a single supergiant star is unlikely to be detected in any survey that uses size information to identify cluster candidates \citep{Silva-Villa2011}. 

At the high-mass end, the main difficulty is the low birthrates, resulting in poor statistics. Although a large galaxy like the Milky Way may have formed hundreds of clusters with $M>10^5 M_\odot$ over its lifetime (see below),  the numbers that are accessible to detailed study are generally much less than that. Ground-based imaging is generally restricted to ages $<10^8-10^9$ years, when the clusters are bright and blue enough to be separated from foreground stars \citep{Larsen2000a,Dowell2008}. HST imaging can use size information to avoid confusion with individual stars, but typically covers only a small fraction of nearby galaxies.

\subsection{Hidden massive clusters in the Milky Way?}
\label{sec:mw}

If the total star formation rate in (long-lived) clusters and the shape of the ICMF are known, we can calculate how many clusters in a given mass range should have formed over some time interval. Observationally, it is challenging to constrain the current formation rate of star clusters in the Milky Way. To get an order-of-magnitude estimate, we assume a total star formation rate of the Milky Way of about $1 M_\odot$ yr$^{-1}$ \citep{Robitaille2010} and that about 10\% of this is in the form of bound star clusters \citep{Lada2003,Bastian2008,Silva-Villa2010}. In order to estimate the expected number of massive clusters for a pure $M^{-2}$ ICMF, it is necessary to adopt upper and lower limits. Plausible values may be $\sim10 M_\odot$ and $\sim10^6 M_\odot$.  In this case, a cluster formation rate of $0.1 M_\odot$ yr$^{-1}$ yields mean birthrates of $\sim0.9$ Myr$^{-1}$ for $M>10^4 M_\odot$ and $\sim0.08$ Myr$^{-1}$ for $M>10^5 M_\odot$. If the star formation rate was constant in the past, the total number of clusters formed over the past 10 Gyr would then be $\sim8000$ and $\sim800$ for $M>10^4 M_\odot$ and $M>10^5 M_\odot$, respectively. These are likely lower limits, as the SFR may well have been higher in the past, although many of these clusters will have dissolved. However, the fact that only $\sim$one cluster with $M\sim10^5 M_\odot$ is expected to form every 10 Myr may be consistent with the observation that even the most massive young clusters known in our Galaxy barely reach $10^5 M_\odot$ and all are only a few Myr old. Due to complicated and poorly quantified selection effects, the Milky Way may not be the best place for constraining the ICMF at high masses. 

\subsection{Constraints from other galaxies}

\begin{figure}
\center
\includegraphics[width=9cm]{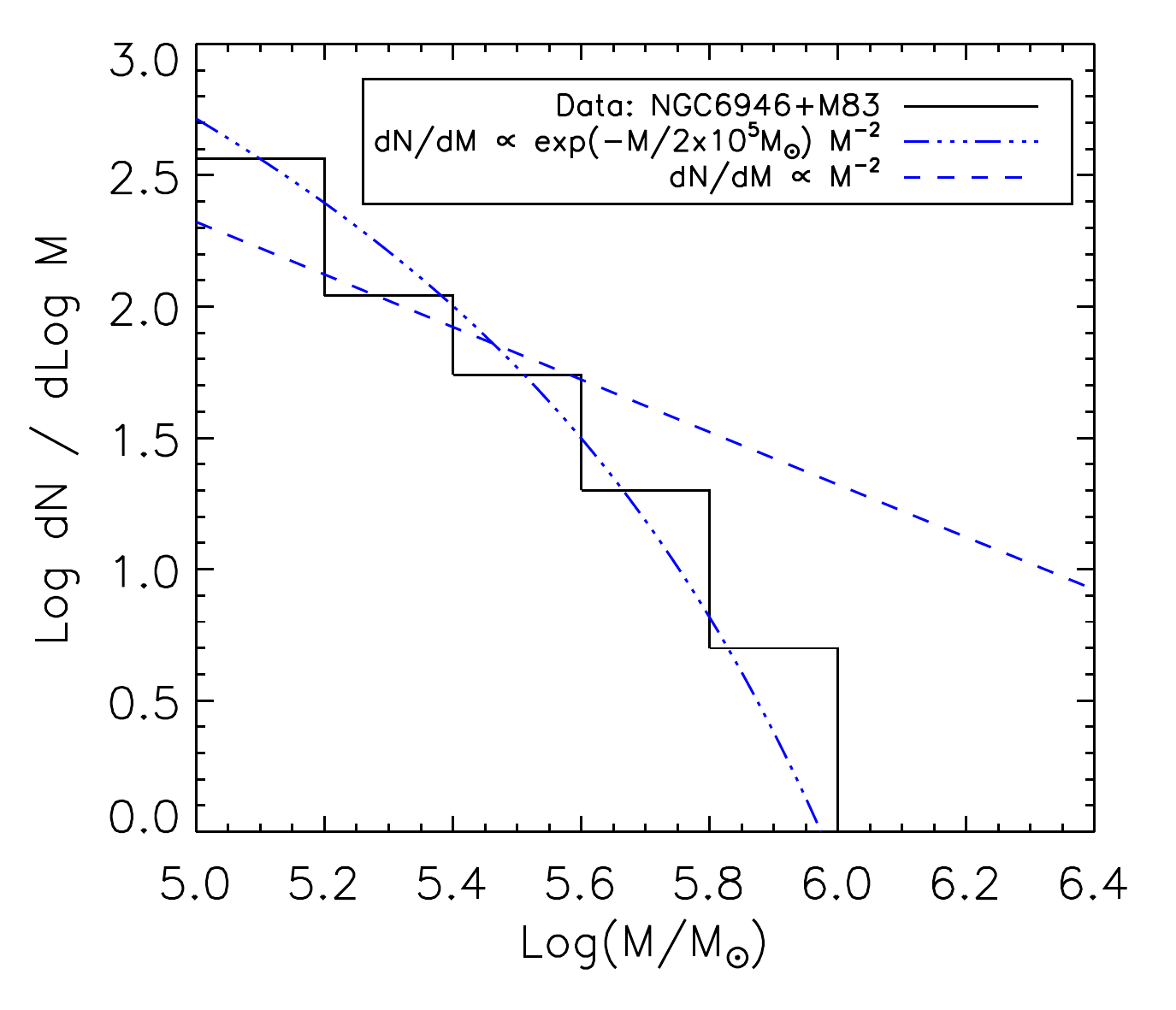} 
\caption{\label{fig:spirals} Observed mass function for the combined cluster populations of NGC~6946 and M83 for clusters younger than 200 Myr. The dashed line shows an $M^{-2}$ power-law while the dotted-dashed line is a Schechter function with a cut-off mass at $2\times10^5 M_\odot$.
}
\end{figure}

As noted above, a number of studies have attempted to constrain the shape of the ICMF in several external galaxies, often finding it to be well represented by a $dN/dM \propto M^{-2}$ shape \emph{over some mass range}. However, many of these studies have been based on HST pointings covering only a small fraction of the target galaxies, or have been targetting galaxies that are relatively small in the first place (like the LMC) and the behaviour of the ICMF is typically poorly constrained above $\sim10^5 M_\odot$. In an attempt to obtain better statistics at high masses, \citet{Larsen2009} used the ground-based imaging of \citet{Larsen1999a} to study the MF of young star clusters in a sample of spiral galaxies. Figure~\ref{fig:spirals} shows the resulting combined mass function for the two spiral galaxies NGC~5236 and NGC~6946, both of which have high current star formation rates and rich populations of young star clusters. The figure includes objects younger than 200 Myr, where the data are still reasonably complete for $M > 10^5 M_\odot$.
The MF is here sampled up to $\sim10^6 M_\odot$, and is poorly fit by a pure $dN/dM\propto M^{-2}$ power-law over the range $10^5 - 10^6 M_\odot$ (dashed line). Instead, a Schechter mass function with a cut-off mass of $2\times10^5 M_\odot$, i.e., $dN/dM \propto \exp(-M/2\times10^5 M_\odot) M^{-2}$, provides an excellent fit to the data (dotted-dashed line). There is no particular physical reason for choosing the Schechter function as a fitting formula, except that it preserves the $M^{-2}$ shape found at lower masses, and provides a convenient way to parameterize a steepening or ``cut-off'' at higher masses.

The data in Fig.~\ref{fig:spirals} allow us to make an independent estimate of the number of massive clusters expected in the Milky Way. The combined star formation rate of NGC~6946 and NGC~5236 is about 5 $M_\odot$ yr$^{-1}$ and the sample contains 111 clusters with $M > 10^5 M_\odot$ and ages younger than 200 Myr. Scaling by the SFR, we would thus expect about 20 clusters in the same mass- and age range in the Milky Way. This is in surprisingly good agreement with the birthrate of about one cluster with $M > 10^5 M_\odot$ every 10 Myr estimated above (Sect.~\ref{sec:mw}).

Although an attractive attribute of the ground-based data is that they cover the full galaxies, there is clearly some risk that cluster samples may be contaminated by more extended groupings of stars. Although it appears unlikely that  contamination would lead to what is essentially a \emph{dearth} of high-mass objects compared to extrapolation of a pure $M^{-2}$ power-law, the data shown in Fig.~\ref{fig:spirals} should still be treated with a healthy dose of skepticism. 
Using HST imaging of NGC~5236, \citet{Chandar2010a} find the MF to be consistent with a cut-off above $10^5 M_\odot$, while \citet{Chandar2011} find cut-offs of $M^\star > 2\times10^5 M_\odot$ to be consistent with observations of NGC~5194. Hence, these studies are essentially consistent with a cut-off mass of several $10^5 M_\odot$.

\begin{figure}
\center
\includegraphics[width=9cm]{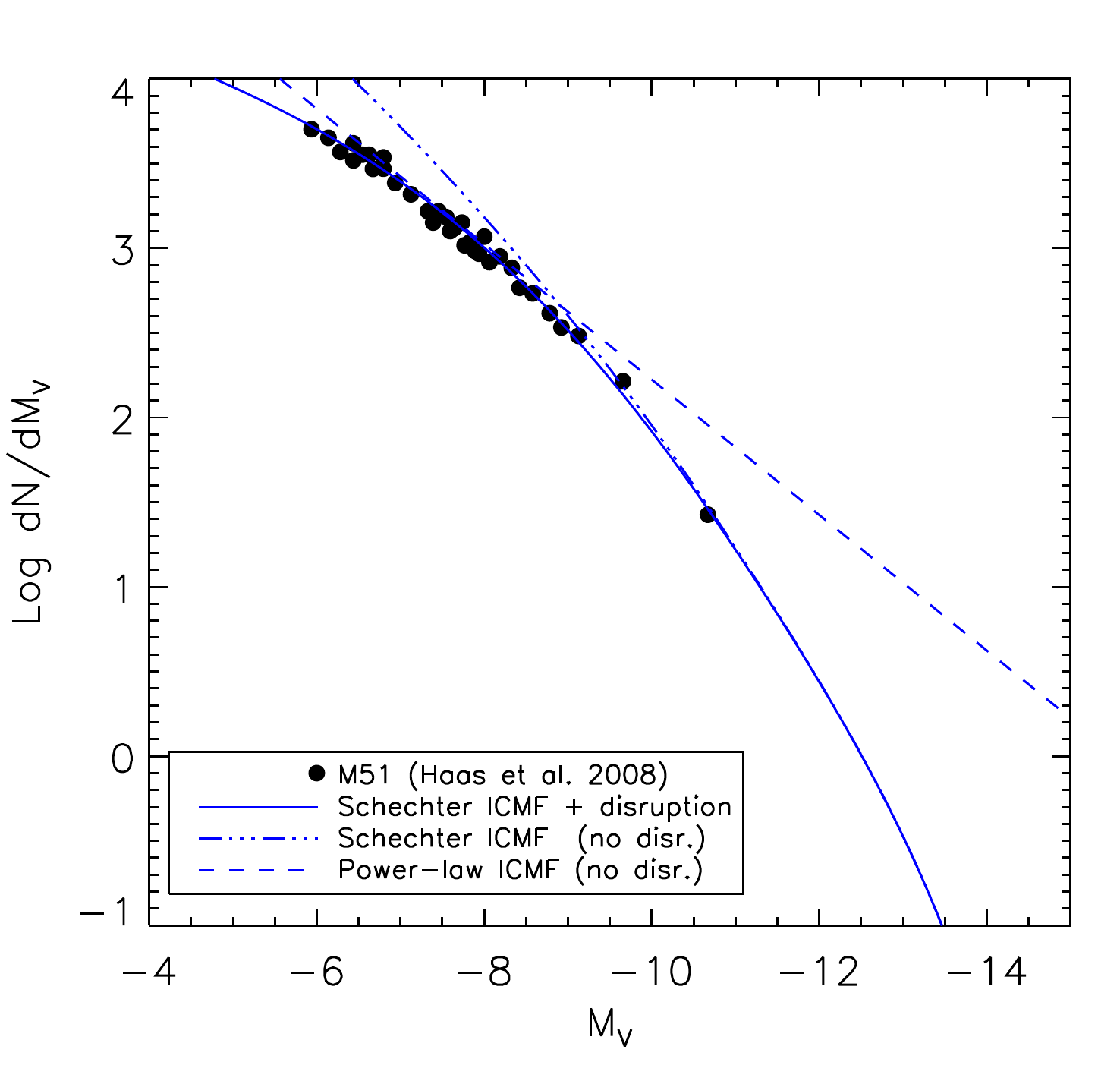} 
\caption{\label{fig:lf_m51} Dots: Observed luminosity function for clusters in M51 = NGC~5194. The dotted line is an $L^{-2}$ power-law, while the solid and dotted-dashed curves are LFs computed for an underlying Schechter ICMF with and without disruption.
}
\end{figure}

A more indirect way to constrain the ICMF is by means of the cluster \emph{luminosity} function (LF). Although the LF is related to the ICMF, it is in general not possible to directly determine the ICMF from observations of the LF as this would require a detailed knowledge of the cluster formation history and dissolution. However, the LF can be predicted for various scenarios and then compared with observations. As an example, Fig.~\ref{fig:lf_m51} shows the observed luminosity function for young clusters in NGC~5194, based on a large HST/ACS mosaic covering essentially the complete galaxy \citep{Haas2008}. A very similar LF, based on the same data, has been shown by \citet{Hwang2008}. Also shown in the figure are predicted LFs for various scenarios. The dashed line shows the LF for a pure power-law ICMF ($dN/dM_i \propto M_i^{-2}$) and no disruption. In this case, the LF itself will have a similar form, $dN/dL \propto L^{-2}$, which is not consistent with the data that instead show a considerably steeper slope at the bright end \citep[$dN/dL \propto L^{-2.5}$ for $M_V < -8$;][]{Haas2008}. Instead, a model LF based on a Schechter ICMF with cut-off mass $M^\star = 2\times10^5 M_\odot$ provides a good fit to the data, especially if some mass-dependent disruption is included. We have here assumed that cluster lifetimes depend on mass as $t_{\rm dis} = 5\times10^8 {\rm yr} (M / 10^4 M_\odot)^{0.65}$.

Although luminosity functions are a somewhat indirect way to constrain the ICMF, they are observationally straight forward to determine and do not depend on model-dependent conversions from the observables to ages and masses. It is therefore not surprising that LFs are available for more cluster systems, compared to the relatively small number of systems where MFs have been derived. It is easy to show that a uniform $M^{-2}$ MF will also yield an $L^{-2}$ LF, if there is no disruption.
Unless high-mass clusters are preferentially disrupted, an initial $M^{-2}$ power-law ICMF should therefore always lead to a LF with the same $-2$ slope or shallower. Instead, most studies tend to find LFs that are \emph{steeper} than this, with slopes getting increasingly steeper if they are measured at brighter magnitudes \citep{Larsen2002a,Gieles2010a}.

 Both direct ICMF determinations and LF measurements appear to be consistent with a cut-off mass $\sim2\times10^5 M_\odot$ in normal spiral discs \citep{PortegiesZwart2010}, although there is a hint that the cut-off may be somewhat radially dependent in NGC~5236 (Bastian, these proceedings). The cut-off mass does \emph{not} appear to be universal but must be significant higher in starburst galaxies, which often host large numbers of clusters with masses of $10^6 M_\odot$ or above. For the best-studied on-going major merger, the Antennae, \citet{Zhang1999} suggested a truncation near $10^6 M_\odot$ while a formal fit to their data yields $M_C\sim8\times10^5 M_\odot$ \citep{Jordan2007}.
In old GC systems there also appears to be an upper cut-off mass near $\sim10^6 M_\odot$, with a significant correlation with host galaxy luminosity \citep{Jordan2007}. This may suggest that GCs formed under conditions that were more similar to those in present-day starbursts than quiescent discs.

\section{The internal star formation histories of clusters}
\label{sec:cmds}

The preceding discussion does not suggest a natural division between globular and other clusters based on mass alone. Clusters are sampled from a continuous mass distribution that appears to be well approximated by a Schechter function. While the cut-off mass probably depends on environment, it is on the order of $10^5 M_\odot$ even in present-day star forming disc galaxies. Likewise, the sizes of young massive star clusters both in spiral galaxies and in starbursts/mergers appear similar to those of ancient globular clusters, with typical half-light radii of a few (2--4) pc \citep[e.g.][]{Whitmore1999,Larsen2004}.

Of course, there are potentially other parameters than size and mass to consider. In the Milky Way, most open clusters are (more or less per definition) associated with the disc, and tend to have roughly Solar metallicity \citep{Friel2002}. Most globular clusters, by contrast, have metallicities well below solar and can be grouped into two populations with mean overall metallicities [Fe/H]$\sim-1.5$ and [Fe/H]$\sim-0.5$ \citep[e.g.][]{Zinn1985}, although GCs in giant elliptical galaxies tend to reach near-solar metallicities \citep{Peng2006}. GCs also differ from open clusters by typically having enhanced $\alpha$-element abundances (relative to Fe). However, these differences mostly reflect the properties of the stellar populations with which the open and globular clusters are associated, and should not necessarily be taken as indications of fundamentally different formation mechanisms.

A perhaps more important clue comes from the degree of \emph{internal} homogeneity of the clusters. Open clusters are generally very homogeneous, with no significant age spreads or star-to-star differences in chemical composition \citep[e.g.][]{Pancino2010}. Globular clusters, instead, are well known for exhibiting internal variations in the abundances of some light elements \citep{Kraft1979}, in particular with anti-correlated abundances of O/Na and Mg/Al \citep{Gratton2001}. This suggests that stars in these clusters formed from material that had been processed to different degrees via the CNO cycle, either in AGB or massive (single or binary) main sequence stars \citep{DErcole2010,Decressin2007,DeMink2009}. In either case, star formation must then have been an extended process (tens to hundreds of Myr), in contrast to the conventional view that star clusters form very quickly \citep[e.g.][]{Elmegreen2000} and data for Milky Way open clusters. Furthermore, high-quality colour-magnitude diagrams have now revealed the presence of multiple main sequences and/or sub-giant brances in several globular clusters, seemingly requiring strongly enhanced He abundances in some of the stars \citep[e.g.][]{Piotto2009,Milone2010}. In summary, globular clusters appear to have experienced complicated star formation- and chemical enrichment histories.

At this point, it is worth recalling that the ancient GCs and present-day open clusters in the galactic disc probe very different regimes of the ICMF. The GCs that have survived until the present day must all have had quite large initial masses, at least $\sim10^5 M_\odot$. Open clusters, on the other hand, are typically low-mass objects with masses less than $10^3 - 10^4 M_\odot$. It is quite possible that the presence of internal abundance variations is linked to the initial cluster mass, with some scenarios predicting that complex internal star formation histories should mainly occur for initial masses greater than $10^5 - 10^6 M_\odot$ \citep{Pflamm-Altenburg2009,Conroy2011}. The real test of whether GCs have a fundamentally different origin than star clusters in the local universe should thus involve objects of similar mass. Unfortunately, little is currently known about the detailed chemical composition of young massive star clusters in the regime $M > 10^5 M_\odot$, simply because most of them are too far away to study individual stars in detail. 

\begin{figure}
\center
\includegraphics[width=15cm]{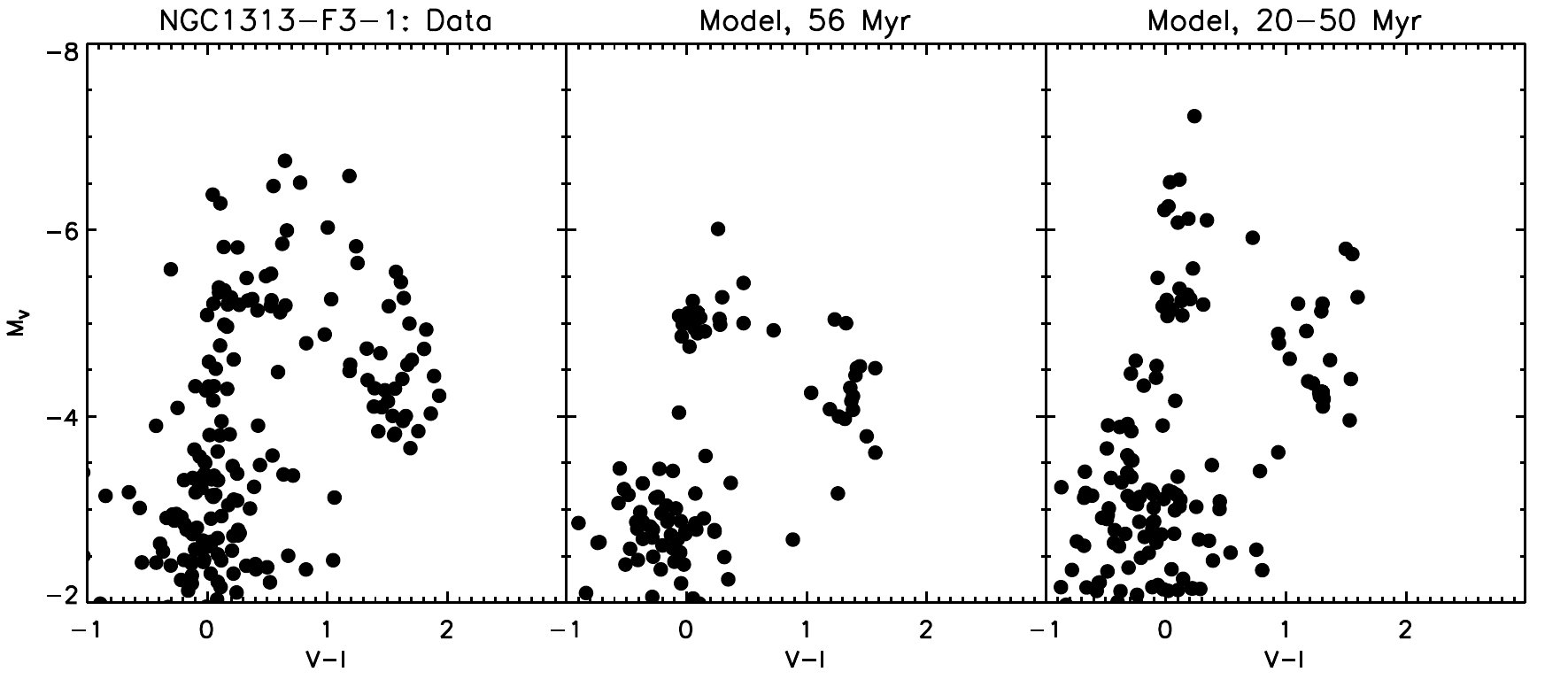} 
\caption{\label{fig:ymccmds} Left: Observed colour-magnitude diagram for the cluster NGC1313-F3-1. Middle: a model based on a single 56 Myr old isochrone. Right: A model including an age range from 20--50 Myr.}
\end{figure}

While abundance anomalies have, so far, only been detected in ancient globular clusters, there is emerging evidence that complex star formation histories may not be unique to old GCs. In particular, age spreads of several 100 Myr have been claimed in some massive, intermediate-age ($\sim2$ Gyr) clusters in the Large Magellanic Cloud \citep{Mackey2008,Milone2009,Girardi2009}, although it has also been suggested that the observed CMD features in these clusters might be at least partially explained by stellar evolutionary effects, e.g. due to stellar rotation \citep{Bastian2009}. 
We have recently presented resolved photometry for a number of younger clusters in several nearby galaxies \citep{Larsen2011}. Figure~\ref{fig:ymccmds} shows one example, for the cluster NGC~1313-F3-1, with an estimated mass of $\sim 2\times10^5 M_\odot$. The left-hand panel shows the photometry, while the centre panel shows a model CMD based on a single isochrone that takes into account observational errors. This model CMD displays a smaller scatter in the luminosities of the supergiant stars, and also exhibits a distinct gap between the main sequence turn-off and the supergiants (the ``blue Hertzsprung gap'', BHG) that is far less clearly seen in the data. The right-hand panel shows a model CMD that includes an age spread between 20 and 50 Myr, which results in a better match to the data. While the multiple-age model is reasonably successful in reproducing the observed CMD for this particular cluster, there are other clusters in the sample of \citet{Larsen2011} for which an age spread cannot, by itself, reproduce the observed CMD features - in particular, the lack of a
BHG seems to be a generic feature of the data, and may also be the result of binary stellar evolution. Mass transfer or mergers in binary systems can produce stars that \emph{appear} to be younger than the actual age of the cluster, and this can produce many of the same effects on the CMD as an age spread.
Thus, the case for multiple stellar populations in younger star clusters remains somewhat ambiguous.

\section{Summary}

Comparison with other nearby spiral galaxies suggests that the Milky Way should contain a significant population of young star clusters with masses above $10^5 M_\odot$. Scaling from the cluster systems of NGC~5236 and NGC~6946, about 20 clusters with $M>10^5 M_\odot$ and ages younger than 200 Myr are expected. Are these objects young analogues of the ancient globular clusters that populate the halo and bulge of our own and other galaxies? In terms of masses and sizes, the young massive clusters in nearby spirals, as well as in on-going starbursts and mergers, certainly appear similar to their older counterparts. Differences in the ``typical'' masses may be largely due to a combination of selection effects and dynamical evolution -- the ``initial cluster mass function'' (ICMF) appears to extend well above $10^5 M_\odot$ whenever there are enough clusters to sample the high-mass end, although there appears to be an environmentally dependent truncation that occurs at a few times $10^5 M_\odot$ in present-day, quiescent discs, but at $>10^6 M_\odot$ in starbursts and ancient GC populations. 

A more significant difference between ancient GCs and present-day cluster formation may be the presence of multiple stellar populations in the former. These are now revealed by direct colour-magnitude diagrams, although the presence of abundance anomalies have hinted at this for many decades. It remains unclear, however, whether these phenomena are uniquely related to ancient GCs. The surviving GCs must have had initial masses well above $10^5 M_\odot$, and little is currently known about the detailed star formation histories of young clusters in this mass range, simply because they are generally too distant to observe individual stars in detail. However, there is evidence from CMDs that some young clusters in the Large Magellanic Cloud, as well as in more distant galaxies, may have extended star formation histories, although this is not entirely conclusive yet. 

%
%
\small  
%
%

%
%
%
%
%

\bibliographystyle{aa}
\bibliography{mnemonic,ref_user}

\end{document}